\documentclass[aps,prb,twocolumn,superscriptaddress,10pt,showkeys]{revtex4-1}

\usepackage{graphicx}
\usepackage{amsmath}
\usepackage{amssymb}
\usepackage{amsfonts}
\usepackage{bm}
\usepackage{bbm}
\usepackage[mathcal]{euscript}
\usepackage{verbatim}

\usepackage{color}

\begin{document}
\title{Extended scheme for the projection of material tensors of arbitrary
symmetry onto a higher symmetry tensor}

\author{Miguel~A. Caro}
\email{mcaroba@gmail.com}
\affiliation{Department of Electrical Engineering and Automation,
Aalto University, Espoo, Finland}
\affiliation{COMP Centre of Excellence in Computational Nanoscience,
Department of Applied Physics, Aalto University, Espoo, Finland}

\newcommand{\eq}[1]{Eq.~(\ref{#1})}
\newcommand{\fig}[1]{Fig.~\ref{#1}}
\newcommand{\ket}[1]{| #1 \rangle}
\newcommand{\bra}[1]{\langle #1 |}
\newcommand{\braket}[2]{\langle #1 | #2 \rangle}

\begin{abstract}
I propose a straightforward
generalization of the projection scheme for elastic tensors introduced by
Moakher and Norris [J. Elasticity \textbf{85}, 215 (2006)]
that takes into account also rotations. The ``closest'' tensor
of any desired symmetry to the original tensor of lower
symmetry is ``closer'' in this generalized scheme.
The method has an important application in the context of the
special quasirandom structure (SQS) method for the computational modeling
of alloys, whereby the supercell's symmetry, and therefore that of the
tensors representing its properties,
is reduced with respect to the material's underlying symmetry.
The approach allows to extract the tensor components most representative
of the macroscopic symmetry of the material.
Although the approach is general, in the present case I apply it to
the elastic tensor and give numerical examples. Simple approximate analytical
expressions for cubic materials are also provided.
\end{abstract}


\date{\today}

\maketitle

In the context of material science, many material properties
are represented by tensors of different ranks. The electric
polarization is given by a rank-1 tensor $P_i$, strain is
represented by a rank-2 tensor $\epsilon_{ij}$, the piezoelectric
response is given by a rank-3 tensor $e_{ijk}$, elasticity
can be described with a rank-4 tensor $C_{ijkl}$, and so on.
The geometrical particularities of the different materials give
rise to the different symmetries of the tensor representing
their properties, as determined by the material's point
group: hexagonal, cubic, etc.
In some cases, these symmetries are ``slightly'' broken, and
materials with an underlying expected symmetry might present
deviations due to imperfections, impurities, alloying effects,
and similar. Within the frame of computational materials science
it is common practice, in part due to computational limits,
to use finite size supercells to represent actual materials.
The special quasirandom structure (SQS) method~\cite{wei_1990}
is a common approach when modeling alloys in the context
of \textit{ab initio} calculations, such as density functional theory
(DFT).~\cite{hohenberg_1964,kohn_1965} The use of
these SQS leads to supercells whose properties slightly deviate
from the ones expected from the crystallographic class of the
macroscopic alloy. These are due to the small deviations of the finite-size
cell with respect to the perfect supercell geometry of the macroscopic
material.
Take as an example the case of the stiffness
tensor of a cubic material which, employing Voigt notation
for the sake of convenience, can be written in matrix form as
\begin{align}
\mathbb{C}^\text{cub} \equiv \left(
\begin{array}{c c c c c c}
C_{11} & C_{12} & C_{12} & 0 & 0 & 0 \\
C_{12} & C_{11} & C_{12} & 0 & 0 & 0 \\
C_{12} & C_{12} & C_{11} & 0 & 0 & 0 \\
0 & 0 & 0 & C_{44} & 0 & 0 \\
0 & 0 & 0 & 0 & C_{44} & 0 \\
0 & 0 & 0 & 0 & 0 & C_{44} 
\end{array}
\right),
\label{01}
\end{align}
where the elements $\mathbb{C}^\text{cub}_{IJ}$ (with uppercase indices
indicating Voigt notation) are the elastic constants of the material.
There are two sets of symmetries applicable to $\mathbb{C}^\text{cub}$.
The first one is the general intrinsic symmetry of the stiffness
tensor upon exchange of $I \leftrightarrow J$. The second one arises from
the specific cubic symmetry of the crystal class, whereby all three
Cartesian directions are equivalent \textit{within this preferential
reference frame} (I shall come back to this later on), and therefore
$C_{12} = C_{13} = C_{23}$, $C_{11} = C_{22} = C_{33}$ and $C_{44} =
C_{55} = C_{66}$.

When using the SQS approach, e.g. to study the elastic properties of
a cubic alloy, one is left using a ``pseudo-cubic'' unit cell which could have
its symmetry reduced to anything lower than cubic, including
triclinic (i.e. no symmetry at all). Tasn\'adi
\textit{et al}.~\cite{tasnadi_2012} used the SQS approach together
with Moakher and Norris's projector scheme,~\cite{moakher_2006} that I shall
discuss in more detail in the following, to study cubic Ti$_{0.5}$Al$_{0.5}$N
alloys. One of the triclinic $4 \times 3 \times 2$ supercells gave the
following stiffness tensor (in GPa):
\begin{align}
\mathbb{C}^\text{tric} \equiv \left(
\begin{array}{c c c c c c}
436 & 161 & 160 & 12 & 11 & 25 \\
 & 453 & 160 & 4 & 15 & 1 \\
 &  & 428 & 13 & 3 & 8 \\
 &  &  & 188 & 12 & 9 \\
\multicolumn{3}{c}{\text{SYM}} & & 186 & 9 \\
 &  &  &  &  & 189 
\end{array}
\right),
\label{02}
\end{align}
which resembles a cubic stiffness tensor [\eq{01}] but strictly speaking has
inherited the triclinic symmetry (no symmetry) of the parent supercell. In
order to extract the cubic ``part'' of \eq{02}, Tasn\'adi \textit{et al.}
resorted to the projection scheme introduced by Moakher and Norris. This
method relies on projector operators $P^\text{sym}$
that project any given tensor,
for instance $\mathbb{C}^\text{tric}$ in \eq{02},
onto the closest tensor of a higher
symmetry of choice $\mathbb{C}^\text{sym}$, by means of minimizing the
Euclidean distance between the two, $\| \mathbb{C}^\text{tric}
- \mathbb{C}^\text{cub} \|$ in this case. The reader is referred
to Ref.~\onlinecite{moakher_2006} for the details of the method.
Moakher and Norris's approach makes for an elegant and powerful formalism.
In the particular case of \eq{02},
the projection onto $\mathbb{C}^\text{cub}$
is done using $P^\text{cub}$ (see
Refs.~\onlinecite{tasnadi_2012,moakher_2006} for its specific form) as
\begin{align}
\mathbb{C}^\text{cub} = P^\text{cub} \mathbb{C}^\text{tric},
\end{align}
which yields average projected cubic elastic constants of
\begin{align}
\bar{C}^\text{cub}_{11} = \frac{C^\text{tric}_{11} + C^\text{tric}_{22} +
C^\text{tric}_{33}}{3} = 439.0 \,\text{GPa},
\nonumber \\
\bar{C}^\text{cub}_{12} = \frac{C^\text{tric}_{12} + C^\text{tric}_{13} +
C^\text{tric}_{23}}{3} = 160.3 \,\text{GPa},
\nonumber \\
\bar{C}^\text{cub}_{44} = \frac{C^\text{tric}_{44} + C^\text{tric}_{55} +
C^\text{tric}_{66}}{3} = 187.7 \,\text{GPa},
\end{align}
and Euclidean distance $\| \mathbb{C}^\text{tric} - \mathbb{C}^\text{cub}
\| = 91.0 \,\text{GPa}$.
The problem with this approach is that it does not take into account the
rotational degrees of freedom. The cubic elastic tensor takes the form
indicated in \eq{02} only if the crystallographic axes coincide with the
Cartesian axes in which the representation of the tensor is carried out.
Because the material properties are not affected by rotations, one can
further reduce the Euclidean distance between $\mathbb{C}^\text{tric}$
and $\mathbb{C}^\text{cub}$ by rotating either one of them so as to
maximize the projection. In fact, Moakher and Norris's projector scheme
implemented \textit{as is} would lead to the inconsistency that a perfectly
cubic elastic tensor, rotated with respect to the reference frame in which
the projector is obtained, would \textit{differ} from its own
cubic projection in that
reference frame. The issue of system orientation was already highlighted in
their original work.~\cite{moakher_2006}
If the triclinic stiffness tensor of \eq{02} is rotated
by $-1.89$, $-1.83$ and $+6.37$ degrees (where ``$+$'' means
counterclockwise and
``$-$'' clockwise) with respect to the first, second and third
Cartesian axes respectively, in that order, then the projection
of $\mathbb{C}^\text{tric}$ onto its closest cubic elastic tensor
yields $\bar{C}^\text{cub}_{11} = 436.8 \,\text{GPa}$,
$\bar{C}^\text{cub}_{12} = 161.4 \,\text{GPa}$ and
$\bar{C}^\text{cub}_{44} = 188.7 \,\text{GPa}$. More importantly, the
Euclidean distance between $\tilde{\mathbb{C}}^\text{tric}$
and $\mathbb{C}^\text{cub}$ is further reduced to 83.7~GPa, where
the tilde denotes the rotation performed. Note that in this case the
difference between the two different approaches is small because
$\mathbb{C}^\text{tric}$ was already ``almost'' cubic (the rotation
angles to correct the structure are correspondingly small). In a more general
case, the input tensor might be in a form that does not closely
resemble the symmetry of interest, leading to even
larger discrepancies. Based on similar considerations, Diner \textit{et al}.
have recently presented a rotation-based method to identify the symmetry
class of a tensor which is not expressed in its natural coordinate
system.~\cite{diner_2011}

Since, as previously mentioned, rotations do not affect the properties
of materials, only their mathematical representation, the procedure
introduced here allows to obtain a better (``closer'') higher-symmetry 
projection of the original tensor. In the following I deal with
the details of the present approach in the context of Moakher and Norris's
method, whose details are given in Ref.~\onlinecite{moakher_2006}.

Let $\mathbb{C}$ and $\mathbb{C}^\text{sym}$ be material tensors of
arbitrary symmetry and specific symmetry ``sym'', respectively. Then
$\mathbb{C}^\text{sym}$ can be expressed as a linear combination of basis
tensors $\mathbb{V}_i$ as~\cite{moakher_2006}
\begin{align}
\mathbb{C}^\text{sym} = \sum\limits_{i=1}^N a_i \mathbb{V}_i,
\end{align}
where $a_i$ are constant coefficients and $N$ is the size of the basis. For
the elasticity of cubic materials, there are 3 independent elastic constants
and therefore $N=3$.

Now, as I have discussed, a rotation of $\mathbb{C}$ does not affect its
properties, only its representation, and thus I can define the general
rotation operator
\begin{align}
R (\theta_x, \theta_y, \theta_z) \equiv R_z (\theta_z) R_y (\theta_y)
R_x (\theta_x).
\end{align}
In matrix form, each individual rotation takes the following form:
\begin{align}
R_x = \left(
\begin{array}{c c c}
1 & 0 & 0 \\
0 & \cos{\theta_x} & -\sin{\theta_x} \\
0 & \sin{\theta_x} & \cos{\theta_x}
\end{array}
\right),
\nonumber \\
R_y = \left(
\begin{array}{c c c}
\cos{\theta_y} & 0 & \sin{\theta_y} \\
0 & 1 & 0 \\
-\sin{\theta_y} & 0 & \cos{\theta_y}
\end{array}
\right),
\nonumber \\
R_z = \left(
\begin{array}{c c c}
\cos{\theta_z} & -\sin{\theta_z} & 0 \\
\sin{\theta_z} & \cos{\theta_z} & 0 \\
0 & 0 & 1
\end{array}
\right).
\end{align}
The different $\theta_i$ give counterclockwise rotation angles around the
Cartesian axes. Note
that the order in which the different rotations are carried out is important
in determining $R$. Therefore the general form of tensor $\mathbb{C}$
is given by
\begin{align}
\tilde{\mathbb{C}} (\theta_x, \theta_y, \theta_z)  \equiv
R (\theta_x, \theta_y, \theta_z) \mathbb{C}.
\end{align}
In the particular case of the stiffness tensor (rank-4), the
matrix elements of $\tilde{\mathbb{C}}$ are given by
\begin{align}
\tilde{\mathbb{C}}_{ijkl} = \sum\limits_{m,n,o,p} R_{im} R_{jn} R_{ko}
R_{lp} C_{mnop},
\end{align}
where all the indices run from 1 to 3.
I could have equivalently made the rotation operator act on the different
basis components
$\mathbb{V}_i$, however this would lead to complicated angle-dependent
projectors. It seems best and simplest to keep Moakher and Norris's original
formulation for the projectors and introduce the angular dependence on
$\mathbb{C}$ instead. The projector (see Ref.~\onlinecite{moakher_2006})
for a given symmetry sym links
$\tilde{\mathbb{C}}$ and $\mathbb{C}^\text{sym}$ as
\begin{align}
\mathbb{C}^\text{sym} (\theta_x, \theta_y, \theta_z) = P^\text{sym}
\tilde{\mathbb{C}} (\theta_x, \theta_y, \theta_z).
\end{align}

The condition that the Euclidean distance between
$\tilde{\mathbb{C}}$ and $\mathbb{C}^\text{sym}$ be minimized is given
by
\begin{align}
\frac{\partial}{\partial a_i} \| \tilde{\mathbb{C}} (\theta_x, \theta_y,
\theta_z) - \mathbb{C}^\text{sym} (\theta_x, \theta_y, \theta_z) \|^2 = 0,
\label{03}
\end{align}
which is the original condition in the formulation of Moakher and
Norris,~\cite{moakher_2006} plus a new requirement for the rotation angles:
\begin{align}
\frac{\partial}{\partial \theta_i} \| \tilde{\mathbb{C}} (\theta_x, \theta_y,
\theta_z) - \mathbb{C}^\text{sym} (\theta_x, \theta_y, \theta_z) \|^2 = 0.
\label{04}
\end{align}
With the first condition only, \eq{03}, the projector can be obtained
analytically from the $\mathbb{V}_i$ alone,
independent of $\tilde{\mathbb{C}}$, as done by Moakher and Norris. This
can then be fed into \eq{04} and the quantity
\begin{align}
\| \tilde{\mathbb{C}} (\theta_x, \theta_y,
\theta_z) - P^\text{sym} \tilde{\mathbb{C}} (\theta_x, \theta_y,
\theta_z) \|^2
\end{align}
minimized numerically with respect to the rotation angles. Alternatively,
it can be shown that this minimization is equivalent to the condition
of maximum for the Euclidean norm of the projected tensor, $\| P^\text{sym}
\tilde{\mathbb{C}} (\theta_x, \theta_y, \theta_z) \|$, taking the
rotation angles as variational parameters. To carry out this task
in the previous example,
I have used the \texttt{Mathematica} symbolic calculator, using the same
script as given in the Appendix. The procedure is straightforward and can be
readily extended to other computational tools.

\begin{figure}[t]
\includegraphics{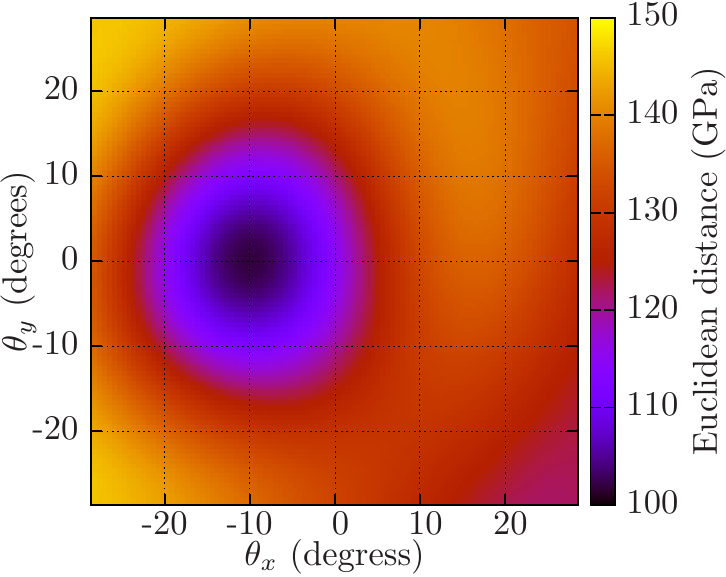}
\caption{(Color online) Euclidean distance between a triclinic elastic tensor
which resembles a hexagonal one, and its hexagonal projection, as a
function of rotation angles about the $x$ and $y$ axes.}
\label{06}
\end{figure}

Note that if an isotropy plane is present then one of the rotation angles
becomes redundant. For example, a hexagonal projection which takes the
$c$ axis as parallel to $z$ maintains a constant Euclidean distance
with the original triclinic tensor for any arbitrary value of $\theta_z$.
This allows to effectively visualize the effect of the present approach
by plotting the Euclidean distance for a hexagonal projection as a
function of $\theta_x$ and $\theta_y$. In order to do this I have
generated the following triclinic elastic tensor by adding a
random~\footnote{True random numbers have been obtained from www.random.org.}
amount in the range between $-25$ and $+25$~GPa to each component of
the perfect
hexagonal elastic tensor of wurtzite GaN (in GPa):~\cite{caro_2012c}
\begin{align}
\mathbb{C} \equiv \left(
\begin{array}{c c c c c c}
352 & 155 & 85 & 2 & -19 & -8 \\
 & 378 & 94 & 13 & -8 & -25 \\
 &  & 395 & -24 & -18 & -19 \\
 &  &  & 103 & 0 & 5 \\
\multicolumn{3}{c}{\text{SYM}} & & 111 & 15 \\
 &  &  &  &  & 118
\end{array}
\right).
\label{05}
\end{align}
Figure~\ref{06} shows a plot of the Euclidean distance between the triclinic
elastic tensor given in \eq{05} and its hexagonal projection as a function
of $\theta_x$ and $\theta_y$, which as discussed is independent of
$\theta_z$. It can be observed that a careless projection without considering
the rotational degrees of freedom would lead to unoptimized calculated
elastic constants, projected at $\theta_x = \theta_y = 0$.
The optimal hexagonal projection requires a previous rotation corresponding
to approximately $\theta_x = -10^\circ$, $\theta_y = 0$.

In the limit where the triclinic tensor resembles the symmetry expected,
for instance cubic in the case of \eq{02}, one can make a number of
assumptions that allow to obtain analytical expressions for the
projected elastic constants. In particular, one can make the assumptions
that i) the components of the rotated triclinic tensor are linear in the
rotation angles ($\sin{x} \approx x$ and $\cos{x} \approx 1$ for small $x$),
ii) that the off-diagonal components of the triclinic tensor are
small compared to the block-diagonal ones (e.g. $C_{16} \ll C_{13}$, etc.)
and iii) that the difference
between the components that should be equal by symmetry for the projected
tensor is small, for instance $C_{11}-C_{33} \ll C_{11}$, etc. in the cubic
case. For the cubic crystal
class, the resulting approximate expressions are:
\begin{align}
\theta_x^\text{cub} \approx \, & \frac{C_{34} - C_{24}}{D^\text{cub}},
\qquad
\theta_y^\text{cub} \approx \frac{C_{15} - C_{35}}{D^\text{cub}},
\nonumber \\
\theta_z^\text{cub} \approx \, & \frac{C_{26} - C_{16}}{D^\text{cub}},
\nonumber \\
C_{11}^\text{cub} \approx \, & \frac{1}{3} \left( C_{11} + C_{22} + C_{33} \right)
+ \frac{4}{3} E^\text{cub},
\nonumber \\
C_{12}^\text{cub}\approx \, & \frac{1}{3} \left( C_{12} + C_{13} + C_{23} \right)
- \frac{2}{3} E^\text{cub},
\nonumber \\
C_{44}^\text{cub} \approx \, & \frac{1}{3} \left( C_{44} + C_{55} + C_{66} \right)
- \frac{2}{3} E^\text{cub},
\end{align}
with
\begin{align}
D^\text{cub} = \, & \frac{2}{3}\left( C_{11} + C_{22} + C_{33} \right) -
\frac{2}{3} \left( C_{12} + C_{13} + C_{23} \right)
\nonumber \\
& - \frac{4}{3} \left( C_{44} + C_{55} + C_{66} \right),
\nonumber \\
E^\text{cub} = \, & \left( C_{34} - C_{24} \right) \theta_x^\text{cub}
+ \left( C_{15} - C_{35} \right) \theta_y^\text{cub}
\nonumber \\
& + \left( C_{26} - C_{16} \right) \theta_z^\text{cub}.
\end{align}
Equivalent expressions for other crystal classes are considerably
lengthier than for the cubic one, and therefore I have opted to
only report the latter as way of example.
Obviously, the numerical solution is always more accurate, in particular
when large rotation angles are required and/or the tensor does not closely
resemble the target projection symmetry.

In summary, I have presented an extension to Moakher and Norris's
formalism~\cite{moakher_2006}
that allows to find the material tensor closest to a tensor of lower symmetry
including also the rotational degrees of freedom. These rotational
degrees of freedom do not determine
the symmetry properties of the tensor but rather their mathematical
representation and therefore ought to be considered when searching
for higher symmetry representations of the tensor's properties. I have
explicitly
carried out this calculation for the stiffness tensor and exemplified it
in the context of the SQS approach, for which these considerations
are readily applicable. Approximate analytical expressions have
been provided for the case of the cubic crystal class.

The author is thankful to R\'{e}mi Zoubkoff for the critical reading
of this manuscript.

\appendix*

\section{Script for angle optimization}

The following \texttt{Mathematica} script performs the
angle-optimized projection of the triclinic elastic tensor given in \eq{02}
leading to the calculation of the projected elastic constants and
rotation angles given throughout the text.

\begin{widetext}
\footnotesize
\begin{verbatim}
(* Declare triclinic elastic tensor and empty rotated triclinic tensor *)

ctensor = Array[ct, {3, 3, 3, 3}];
rotctensor = Array[0 &, {3, 3, 3, 3}];

(* Define rotation matrices *)

Rx = {{1, 0, 0}, {0, Cos[tx], -Sin[tx]}, {0, Sin[tx], Cos[tx]}};
Ry = {{Cos[ty], 0, Sin[ty]}, {0, 1, 0}, {-Sin[ty], 0, Cos[ty]}};
Rz = {{Cos[tz], -Sin[tz], 0}, {Sin[tz], Cos[tz], 0}, {0, 0, 1}};
R = Rz.Ry.Rx;

(* Rotate triclinic elastic tensor *)

Do[rotctensor[[i, j, k, l]] = Sum[R[[i, m]]*R[[j, n]]*R[[k, o]]*R[[l, p]]*ctensor[[m, n, o, p]],
   {m, 1, 3}, {n, 1, 3}, {o, 1, 3}, {p, 1, 3}], {i, 1, 3}, {j, 1, 3}, {k, 1, 3}, {l, 1, 3}];

(* Assign numerical values to the elements of the triclinic elastic tensor. Note symmetries applied *)

ct[1, 1, 1, 1] = 436; ct[2, 2, 2, 2] = 453; ct[3, 3, 3, 3] = 428;
ct[1, 1, 2, 2] = ct[2, 2, 1, 1] = 161;
ct[1, 1, 3, 3] = ct[3, 3, 1, 1] = 160; 
ct[2, 2, 3, 3] = ct[3, 3, 2, 2] = 160;
ct[1, 2, 1, 2] = ct[1, 2, 2, 1] = ct[2, 1, 1, 2] = ct[2, 1, 2, 1] = 189;
ct[1, 3, 1, 3] = ct[1, 3, 3, 1] = ct[3, 1, 1, 3] = ct[3, 1, 3, 1] = 186;
ct[2, 3, 2, 3] = ct[2, 3, 3, 2] = ct[3, 2, 2, 3] = ct[3, 2, 3, 2] = 188;
ct[1, 1, 1, 2] = ct[1, 1, 2, 1] = ct[1, 2, 1, 1] = ct[2, 1, 1, 1] = 25;
ct[1, 1, 1, 3] = ct[1, 1, 3, 1] = ct[1, 3, 1, 1] = ct[3, 1, 1, 1] = 11;
ct[1, 1, 2, 3] = ct[1, 1, 3, 2] = ct[2, 3, 1, 1] = ct[3, 2, 1, 1] = 12;
ct[2, 2, 2, 1] = ct[2, 2, 1, 2] = ct[2, 1, 2, 2] = ct[1, 2, 2, 2] = 1;
ct[2, 2, 2, 3] = ct[2, 2, 3, 2] = ct[2, 3, 2, 2] = ct[3, 2, 2, 2] = 4;
ct[2, 2, 1, 3] = ct[2, 2, 3, 1] = ct[1, 3, 2, 2] = ct[3, 1, 2, 2] = 15;
ct[3, 3, 3, 2] = ct[3, 3, 2, 3] = ct[3, 2, 3, 3] = ct[2, 3, 3, 3] = 13;
ct[3, 3, 3, 1] = ct[3, 3, 1, 3] = ct[3, 1, 3, 3] = ct[1, 3, 3, 3] = 3;
ct[3, 3, 2, 1] = ct[3, 3, 1, 2] = ct[2, 1, 3, 3] = ct[1, 2, 3, 3] = 8;
ct[1, 2, 1, 3] = ct[1, 2, 3, 1] = ct[2, 1, 1, 3] = ct[2, 1, 3, 1]
               = ct[1, 3, 1, 2] = ct[1, 3, 2, 1] = ct[3, 1, 1, 2] = ct[3, 1, 2, 1] = 9;
ct[2, 1, 2, 3] = ct[2, 1, 3, 2] = ct[1, 2, 2, 3] = ct[1, 2, 3, 2]
               = ct[2, 3, 2, 1] = ct[2, 3, 1, 2] = ct[3, 2, 2, 1] = ct[3, 2, 1, 2] = 9;
ct[3, 2, 3, 1] = ct[3, 2, 1, 3] = ct[2, 3, 3, 1] = ct[2, 3, 1, 3]
               = ct[3, 1, 3, 2] = ct[3, 1, 2, 3] = ct[1, 3, 3, 2] = ct[1, 3, 2, 3] = 12;

(* Express the rotated triclinic elastic tensor in vector form, with 21 components *)
(* Note the Sqrt[2], 2 and 2 Sqrt[2] factors to preserve the norm *)

rotcvector = {rotctensor[[1, 1, 1, 1]], rotctensor[[2, 2, 2, 2]], rotctensor[[3, 3, 3, 3]],
              Sqrt[2] rotctensor[[2, 2, 3, 3]], Sqrt[2] rotctensor[[1, 1, 3, 3]],
              Sqrt[2] rotctensor[[1, 1, 2, 2]], 2 rotctensor[[2, 3, 2, 3]], 2 rotctensor[[1, 3, 1, 3]],
              2 rotctensor[[1, 2, 1, 2]], 2 rotctensor[[1, 1, 2, 3]], 2 rotctensor[[2, 2, 1, 3]],
              2 rotctensor[[3, 3, 1, 2]], 2 rotctensor[[3, 3, 2, 3]], 2 rotctensor[[1, 1, 1, 3]],
              2 rotctensor[[2, 2, 1, 2]], 2 rotctensor[[2, 2, 2, 3]], 2 rotctensor[[3, 3, 1, 3]],
              2 rotctensor[[1, 1, 1, 2]], 2 Sqrt[2] rotctensor[[1, 3, 1, 2]], 2 Sqrt[2] rotctensor[[2, 3, 1, 2]], 
              2 Sqrt[2] rotctensor[[2, 3, 1, 3]]};
              
(* Obtain the cubic projector following Moakher and Norris' recipe *)

(* Define the three basis elements for the cubic elastic tensor in vector form *)

velacub[1] = {1, 1, 1, 0, 0, 0, 0, 0, 0, 0, 0, 0, 0, 0, 0, 0, 0, 0, 0, 0, 0};
velacub[2] = {0, 0, 0, Sqrt[2], Sqrt[2], Sqrt[2], 0, 0, 0, 0, 0, 0, 0, 0, 0, 0, 0, 0, 0, 0, 0};
velacub[3] = {0, 0, 0, 0, 0, 0, 1, 1, 1, 0, 0, 0, 0, 0, 0, 0, 0, 0, 0, 0, 0};

(* Generate projector *)

delacub = Array[0 &, {3, 3}];
Do[delacub[[i, j]] = velacub[i].velacub[j], {i, 1, 3}, {j, 1, 3}]
idelacub = Inverse[delacub];
Pelacub = Sum[idelacub[[i, j]]*Outer[Times, velacub[i], velacub[j]], {i, 1, 3}, {j, 1, 3}];

(* Numerical minimization of the Euclidean distance between the original triclinic tensor and its cubic projection
   with respect to the rotation angles. Optimized angles tx, ty and tz are in radians *)

NMinimize[(rotcvector - Pelacub.rotcvector).(rotcvector -  Pelacub.rotcvector), {tx, ty, tz}]
{6999.66, {tx -> -0.0329499, ty -> -0.0319465, tz -> 0.111128}}

(* Evaluate cubic projection of the rotated tensor for those angles. Gives the 21 components of the elastic tensor *)

N[Pelacub.rotcvector] /. tx -> -0.0329499 /. ty -> -0.0319465 /. tz -> 0.111128
{436.836, 436.836, 436.836, 228.276, 228.276, 228.276, 377.497, 377.497, 377.497,
 0., 0., 0., 0., 0., 0., 0., 0., 0., 0., 0., 0.}
\end{verbatim}
\end{widetext}
When the normalizing factors are taken into account ($\sqrt{2}$ for $C_{12}$
and 2 for $C_{44}$) the elastic constants of the closest cubic projection of
the original triclinic elastic tensor are obtained.

\end{document}